\documentclass[]{aastex631}

\begin{document}

\title{Observation of the Galactic Center PeVatron Beyond 100 TeV with HAWC}

\correspondingauthor{Sohyoun Yun-C\'arcamo}
\email{yunsoh@umd.edu}

\correspondingauthor{Dezhi Huang}
\email{dezhih@umd.edu}

\correspondingauthor{Rishi Babu}
\email{baburish@msu.edu}

\correspondingauthor{Kwok Lung Fan}
\email{klfan@terpmail.umd.edu}

\author{A.~Albert}
\affiliation{Physics Division, Los Alamos National Laboratory, Los Alamos, NM, USA }

\author{R.~Alfaro}
\affiliation{Instituto de F\'{i}sica, Universidad Nacional Aut\'{o}noma de M\'{e}xico, Ciudad de M\'{e}xico, M\'{e}xico }

\author{C.~Alvarez}
\affiliation{Universidad Aut\'{o}noma de Chiapas, Tuxtla Guti\'{e}rrez, Chiapas, M\'{e}xico}

\author{A.~Andr\'{e}s}
\affiliation{Instituto de Astronom\'{i}a, Universidad Nacional Aut\'{o}noma de M\'{e}xico, Ciudad de M\'{e}xico, M\'{e}xico }

\author{J.C.~Arteaga-Vel\'{a}zquez}
\affiliation{Universidad Michoacana de San Nicol\'{a}s de Hidalgo, Morelia, M\'{e}xico }

\author{D.~Avila Rojas}
\affiliation{Instituto de F\'{i}sica, Universidad Nacional Aut\'{o}noma de M\'{e}xico, Ciudad de M\'{e}xico, M\'{e}xico }

\author{H.A.~Ayala Solares}
\affiliation{Department of Physics, Pennsylvania State University, University Park, PA, USA }

\author[0000-0002-5529-6780]{R.~Babu}
\affiliation{Department of Physics and Astronomy, Michigan State University, East Lansing, MI, USA }

\author{E.~Belmont-Moreno}
\affiliation{Instituto de F\'{i}sica, Universidad Nacional Aut\'{o}noma de M\'{e}xico, Ciudad de M\'{e}xico, M\'{e}xico }

\author{A.~Bernal}
\affiliation{Instituto de Astronom\'{i}a, Universidad Nacional Aut\'{o}noma de M\'{e}xico, Ciudad de M\'{e}xico, M\'{e}xico }

% ...



\begin{abstract}
We report an observation of ultra-high energy (UHE) gamma rays from the Galactic Center region, using seven years of data collected by the High-Altitude Water Cherenkov (HAWC) Observatory. The HAWC data are best described as a point-like source (HAWC J1746-2856) with a power-law spectrum ($\mathrm{d}N/\mathrm{d}E=\phi(E/26 \,\text{TeV})^{\gamma}$), where $\gamma=-2.88 \pm 0.15_{\text{stat}} - 0.1_{\text{sys}} $ and $\phi=1.5 \times 10^{-15}$ (TeV cm$^{2}$s)$^{-1}$ $\pm\, 0.3_{\text{stat}}\,^{+0.08_{\text{sys}}}_{-0.13_{\text{sys}}}$ extending from 6 to 114 TeV. We find no evidence of a spectral cutoff up to $100$~TeV using HAWC data. Two known point-like gamma-ray sources are spatially coincident with the HAWC gamma-ray excess: Sgr A$^{*}$ (HESS J1745-290) and the Arc (HESS J1746-285). We subtract the known flux contribution of these point sources from the measured flux of HAWC J1746-2856 to exclude their contamination and show that the excess observed by HAWC remains significant ($>$5$\sigma$) with the spectrum extending to $>$100~TeV. Our result supports that these detected UHE gamma rays can originate via hadronic interaction of PeV cosmic-ray protons with the dense ambient gas and confirms the presence of a proton PeVatron at the Galactic Center.
\end{abstract}



\section{Introduction} \label{sec:intro}

The Galactic sources of cosmic-ray acceleration to petaelectronvolt (PeV) energies---known as PeVatrons---remain unidentified and are still subject to discussion \citep{aharonian2019massive,2021Univ....7..324C,2023PhRvD.107d3002S,2024NatAs...8..425D,2024arXiv240415944F,blasi2013origin,amato2014origin,gabici2019origin}. Previous studies suggest that cosmic rays are actively accelerated in the Galactic Center (GC) region \citep{2016Natur.531..476H}. The arc-minute angular resolution of the Imaging Atmospheric Cherenkov telescopes (IACTs), e.g., the High Energy Stereoscopic System (H.E.S.S.; \citealp{aharonian2006hess,2016Natur.531..476H,abdalla2018characterising}), the Major Atmospheric Gamma-Ray Imaging
Cherenkov (MAGIC) telescopes \citep{acciari2020magic}, and the
Very-Energetic Radiation Imaging Telescope Array System
(VERITAS; \citealp{adams2021veritas}), has allowed them to measure gamma-ray emission up to $\sim$20 TeV from the two point sources of interest in the region: Sgr A$^{*}$ (HESS J1745-290), the supermassive black hole at the center of the Galaxy, and the unidentified source HESS J1746-285, which is spatially coincident with the Galactic radio arc \citep{yusef1987linear,yusef200420}. The observation of the point-like supernova remnant (SNR) G0.9+0.1 \citep{abdalla2018characterising,acciari2020magic,adams2021veritas} and the unidentified extended source HESS J1745-303 \citep{Aharonian_2006} were reported as well about $1^{\circ}$ away from the GC. These IACTs have also observed very-high-energy (VHE) gamma rays from the GC ridge \citep{2016Natur.531..476H,abdalla2018characterising, adams2021veritas, acciari2020magic}. This diffuse emission spatially correlates to the Central Molecular Zone (CMZ) morphology (\citealp{Aharonian_2006}), which is derived from dense gas tracers \citep{tsuboi1999dense}. This correlation suggests a hadronic origin for the observed gamma-ray emission given the severe energy losses via synchrotron emission in the leptonic scenario \citep{aharonian2006hess,2016Natur.531..476H,abdalla2018characterising}.

In this work, we use seven years of data from the HAWC Gamma-Ray Observatory to study the gamma-ray emission from the GC region. Our analysis extends the previous observations to energies $>$100~TeV, which allows the PeV cosmic-ray interaction to be directly probed. We show that the UHE emission observed by HAWC is most likely from the Galactic ridge emission by subtracting the flux contribution from HESS~J1745-290, as reported in \citet{2016Natur.531..476H}, with good agreement in location and spectrum to other observations \citep{adams2021veritas, acciari2020magic,Abe:2023Hn}, and from HESS~J1746-285, as reported in \citet{abdalla2018characterising}. The latter has been observed by several IACTs also with good agreement, but only in \citet{abdalla2018characterising} is the contribution of underlying diffuse emission additionally taken into account for the source HESS J1746-285. The flux contribution of these two sources at energies $>$100~TeV is extremely small. This indicates that these sources do not contribute solely to the origin of the observed VHE gamma rays. Our result provides evidence of a PeVatron at the center of our Galaxy with the first measurement of nearly 100 gamma-ray events with energies $>$100 TeV.

This Letter is organized as follows: Section \ref{sec:hawc} briefly describes the HAWC data set used in this analysis, Sections~\ref{sec:pev} and \ref{sec:disc} present and discuss the results of the analysis, and in Section \ref{section:conc} we present our conclusions.

\section{HAWC data}\label{sec:hawc}

The HAWC Observatory---located on the side of the Sierra Negra volcano in Puebla, Mexico at 4100 m asl---is made up of 300 water Cherenkov detectors \citep{ABEYSEKARA2023168253}. We apply signal-topology-based cuts to reduce the cosmic-ray background (99.9\% of events detected). We recently updated HAWC's reconstruction algorithms (``Pass~5"), improving its effective area, angular resolution, and gamma/hadron separation at the highest energies and zenith angles. With these improvements, HAWC is able to observe the GC, which culminates at 48$^{\circ}$ zenith \citep{albert2024performance}. As a further check, we verified that the results obtained when reconstructing data from the Crab Nebula when it reaches zenith angles greater than $45^{\circ}$ are in agreement with those reported in the study by \citet{albert2024performance}. 

Using 2546 days of HAWC data, we detect gamma-ray emission from the GC region, with a maximum significance of 6.5$\sigma$ above the background. We analyzed the data with the Pass 5 version of the neural network energy estimator \citep{abeysekara2019measurement,albert2024performance} and included off-array events, which are showers whose cores fall off the main array up to 1.5 times its physical area and improve the sensitivity of HAWC to high zenith angles and high energies \citep{albert2024performance}. 

To model the gamma-ray flux from the GC region we employed the HAWC Accelerated Likelihood (HAL) plugin with the Multi-Mission Maximum Likelihood (3ML)\footnote{https://github.com/threeML/threeML} framework \citep{younk2015high,abeysekara2022characterizing}---a forward-folded maximum-likelihood approach \citep{Vianello:201667}---within a rectangular region of interest (ROI) $\pm 3^{\circ}$ in latitude and $\pm 2.5^{\circ}$ in longitude. 

We define our test statistics (TS) as:
\begin{equation}\label{ts}
\text{TS} = 2\ln{\left(\frac{L_{\text{model}}}{L_{\text{bkg}}}\right)}\,,
\end{equation}
where $L_{\text{model}}$ denotes the maximum likelihood from the source model and $L_{\text{bkg}}$ is background only. According to Wilks’ theorem \citep{wilks1938large}, which applies to HAWC data \citep{Abeysekara_2017}, the TS is asymptotically $\chi^{2}$ distributed, with the degrees of freedom equal to the difference in the number of free parameters of the nested models. Thus, under the case of one free parameter, $\sqrt{\text{TS}}$ can be used as a measure of significance, $\sigma$.

The extended-source assumption was tested and no strong preference was found ($\Delta \text{TS}=6.62$ compared to the point-source assumption), hence the simplest model was chosen. From the extended-source fit, we estimated an upper limit (UL) on the source extension (radius) at the 68\% confidence level (CL) for the Gaussian width of the source ($\sim$0.46$^{\circ}$, see dashed circle in Figure \ref{fig:final}(a)). Adding curvature to the spectrum did not significantly improve the test statistic either ($\Delta \text{TS}=0.44$).

\section{Main analysis results}\label{sec:pev}

The best fit to the data is a point source with a simple-power-law spectrum (TS$=49$ for four free parameters---position and spectral parameters):
\begin{equation}\label{power law}
\frac{\text{d}N}{\text{d}E}=\phi \left(\frac{E}{26 \text{ TeV}}\right) ^{\gamma}\,,
\end{equation}
where $\phi$ is the flux normalization at the pivot energy and $\gamma$ is the power-law index. The pivot energy of 26 TeV is calculated such that it minimizes the correlation between the flux normalization and spectral index. We summarize the best-fit parameters of HAWC J1746-2856 in Table  \ref{tab:sys} and include both statistical and systematic uncertainties. The latter account for the contribution of four non-negligible independent systematic uncertainties that were identified in the previous energy-dependent study of the Crab \citep{abeysekara2019measurement} and are estimated by producing instrument response functions (IRFs) with different detector configurations to investigate any potential mis-modeling of the detector. The results were then compared with the standard HAWC analysis and the uncertainties were added in quadrature. Another source of systematic uncertainty in the flux of HAWC J1746-2856 could be emission from background cosmic rays, often referred to as the cosmic-ray sea, which is thought to have a consistent energy density throughout the Galaxy. Locally, above 100 TeV, the energy density of the cosmic-ray spectrum is approximately $3\times 10^{-4}$~eV/cm$^{3}$ \citep{aguilar2015precision}. For the diffuse emission to significantly impact the results, one would need to assume that the cosmic-ray sea’s flux is nearly a factor of ten higher at the Galactic Center. Even assuming an unusual spectral dependence or normalization of the diffusion coefficient within the CMZ, it would be difficult to explain the at least tenfold discrepancy between the reported local spectrum and the CMZ spectrum observed by HAWC. Additionally, the spectral indices of the cosmic-ray sea and HAWC J1746-2856 are not compatible. Thus, we keep the simplest point-source model.

We calculated an UL on the minimum energy at 6 TeV and a lower limit on the maximum energy at 114 TeV, both at the 68\% CL. Above 100 TeV, the significance of the signal is 1.2~$\sigma$. Above 100 TeV, 3474 events passed trigger conditions for reconstruction, from which 98 events passed HAWC gamma/hadron separation cuts. At 100 TeV, the energy resolution is 10\% in $\log_{10}(E/\text{TeV})$ \citep{abeysekara2019measurement} and the hadron retention after gamma/hadron separation cuts is $<1$~\%. To count the events, we used a circular region centered in the best-fit position (see Table~\ref{tab:sys}) with the radius set at the UL on the source extension.

 \begin{table}[h]
    \centering
    \begin{tabular}{cccc}
    \hline
    Parameter estimated& Best fit &Statistical  & Systematic  \\
  & &uncertainties & uncertainties \\
    \hline
    \hline
     RA ($^{\circ}$) & 266.28 &$\pm$ 0.05 & $+0.09,-0.06$  \\
      Dec ($^{\circ}$)&$-28.94$ &$\pm$ 0.04  &$+0.03,-0.02$  \\
      \hline
     Flux norm. ($\phi$) $\times 10^{-15}$ (TeV cm$^{2}$s)$^{-1}$ &1.5 &$\pm$ 0.30$ $&$+ 0.08,-0.13$\\
     Index ($\gamma$)&$-2.88$ &$\pm$ 0.15 &$-0.1$  \\
        \hline
    \end{tabular}
    \caption{Best-fit results for HAWC J1746-2856 with statistical and systematic uncertainties. The spectrum is best described by a simple power law $\mathrm{d}N/\mathrm{d}E=\phi(E/26 \,\text{TeV})^{\gamma}$. See Section \ref{sec:pev} for details. In Galactic coordinates, the best-fit position of HAWC J1746-2856 is $(l, b) =(0.06^{\circ}$, $0.09^{\circ})$.}
    \label{tab:sys}
\end{table}

\begin{figure}
\begin{center}
\resizebox{1.0\textwidth}{!}{%
\begin{tabular}{cc}
\includegraphics[width=0.49\textwidth]{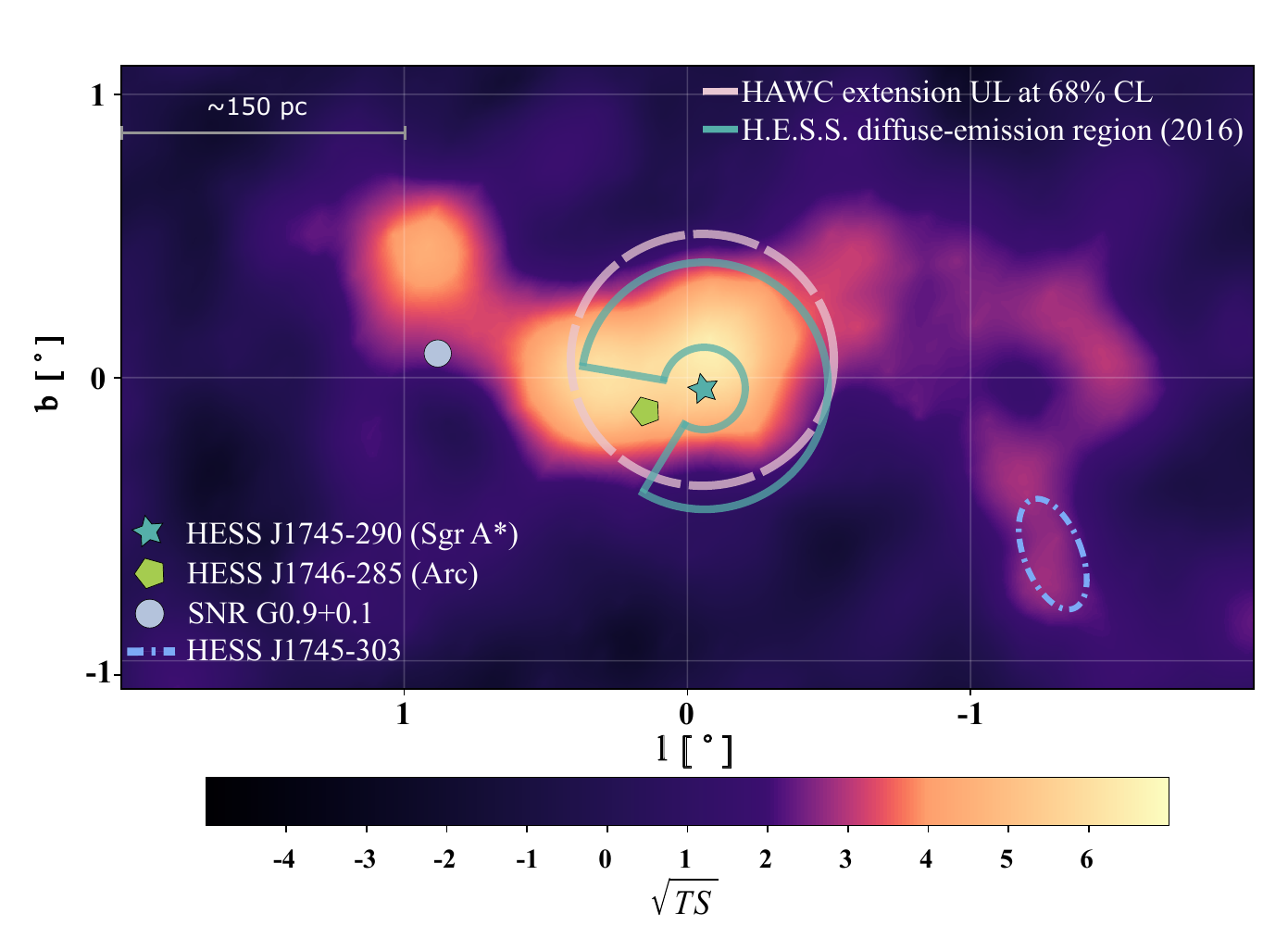}%
&\quad
\includegraphics[width=0.49\textwidth]{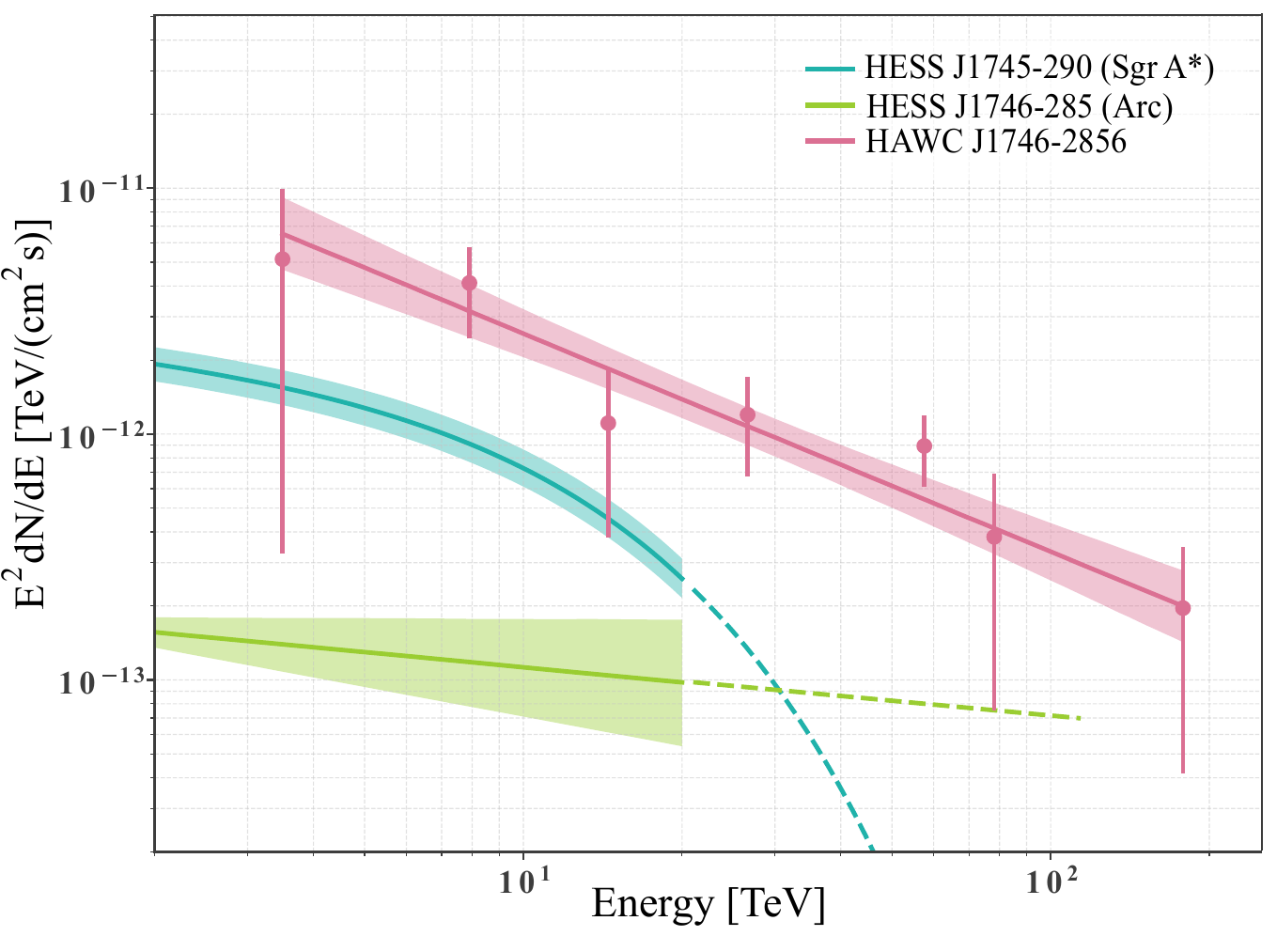}%
\\
(a) & (b)\\
\includegraphics[width=0.49\textwidth]{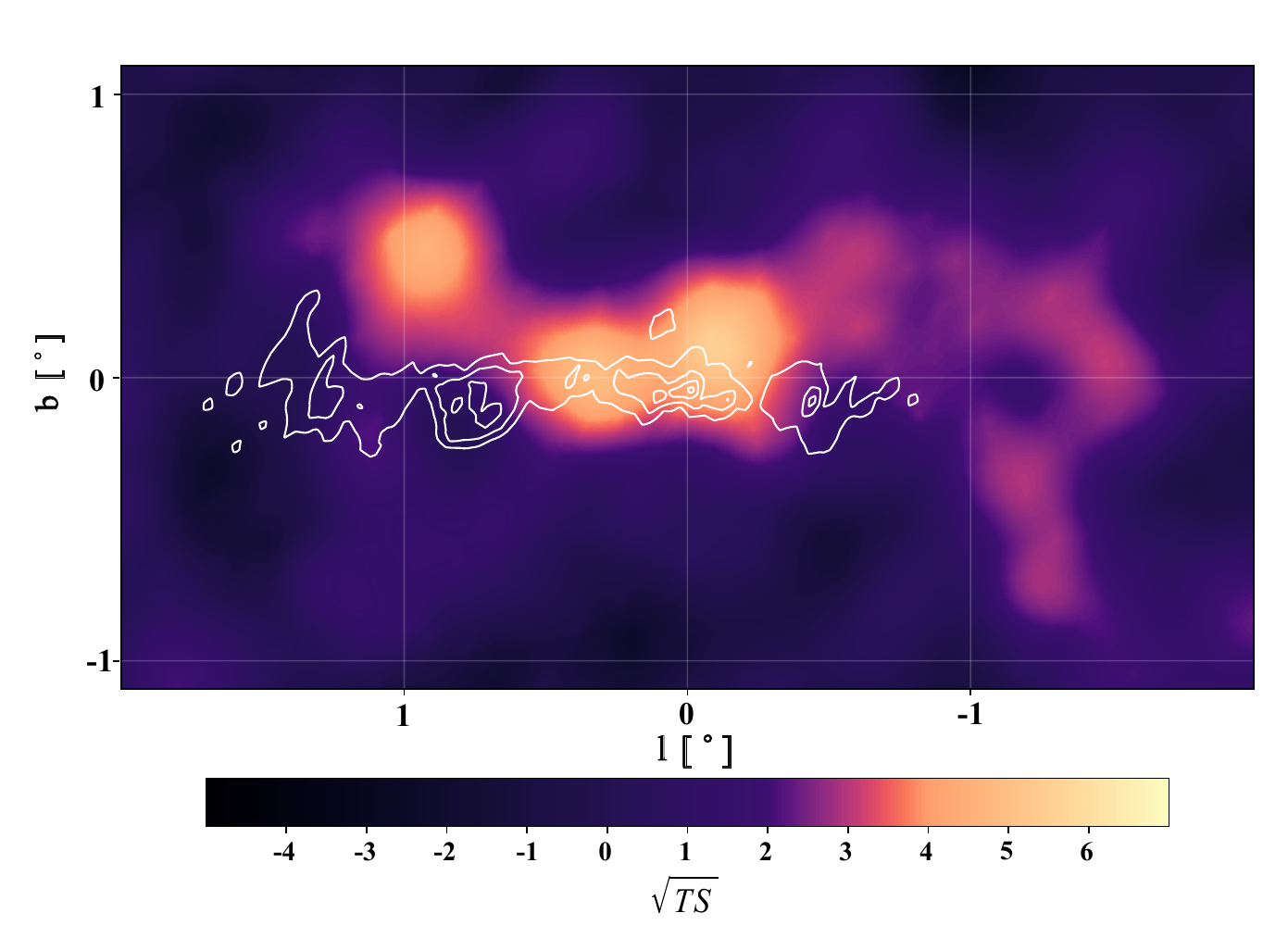}%
&\quad
\includegraphics[width=0.49\textwidth]{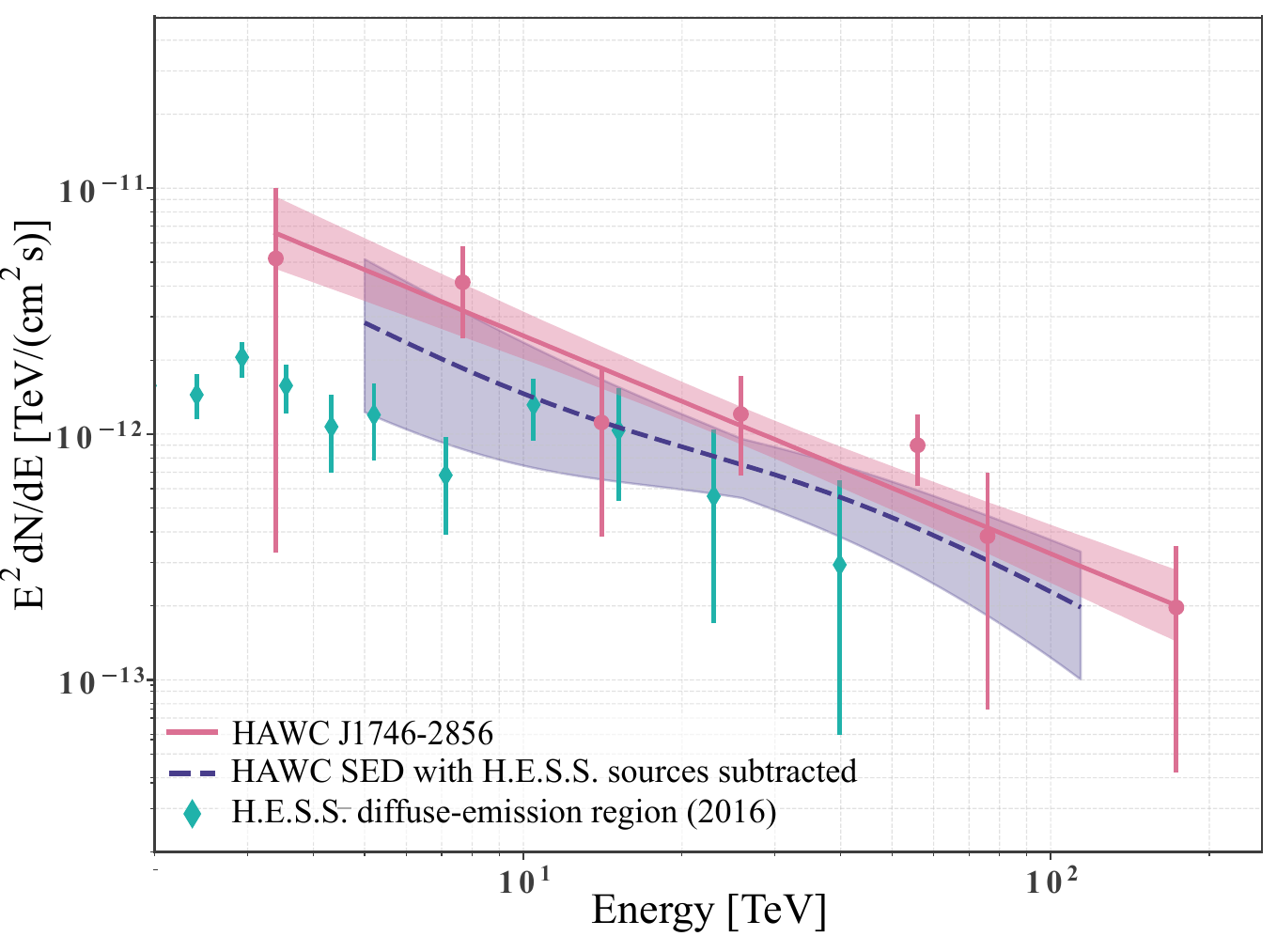}%
\\
(c) & (d) \\
\end{tabular}
}
\caption{Galactic Center analysis results. \textbf{(a)} Significance map obtained using the HAWC neural network energy estimator (on- and off-array events) \citep{abeysekara2019measurement} and the position of the three main point sources and one extended source in the GC region as measured by H.E.S.S.. The dashed circle outlines the extension UL at 68\% CL. We also include the diffuse region used in the H.E.S.S. analysis \citep{2016Natur.531..476H}. \textbf{(b)} Spectra of the two H.E.S.S. sources, along with the best-fit spectrum of HAWC J1746-2856. The dashed lines for the H.E.S.S. sources show the extrapolation of their best-fit to the HAWC energy range. The flux points are calculated for each energy bin \citep{albert2024performance} by fixing all the fit parameters except for the flux normalization. \textbf{(c)} HAWC emission after subtracting the two H.E.S.S. point sources. We also show the density distribution contours of the ambient gas as traced by CS (J1-0) line emission \citep{tsuboi1999dense}. \textbf{(d)} Original best-fit HAWC spectral energy distribution (SED) and the result after subtracting the two H.E.S.S. point-source spectra. As a reference, we include the diffuse emission measured by \citet{2016Natur.531..476H}, as their region is almost spatially coincident with our model. See Section \ref{sec:pev} for details. \label{fig:final}}
\end{center}
\end{figure}

In Figure \ref{fig:final}(a), we show a significance map from the GC region obtained with HAWC data by calculating the TS of every pixel as the ratio of the logarithm of the likelihoods of the signal measured over the expected background \citep{younk2015high,abeysekara2017observation}. We also include the location of HESS J1745-290 (Sgr A$^{*}$;\citealp{Aharonian_2006}) and HESS J1746-285 (the Arc; \citealp{abdalla2018characterising}), which are relevant to this study as they are inside of the HAWC J1746-2856 extension UL radius and excluded from the diffuse emission region used in \citet{2016Natur.531..476H}. In addition, we show the positions of SNR G0.9+0.1 and HESS J1745-303. The reported gamma-ray flux level of the SNR falls below the sensitivity of HAWC at this declination \citep{albert2024performance}. No significant excess is observed by HAWC at the reported location of the SNR or HESS J1745-303. The 4.5$\sigma$ hot spot above the SNR location is not coincident with any known gamma-ray sources, but it aligns with a candidate open stellar cluster \citep{dutra2003ntt}. Although gamma rays are observed in the vicinity of stellar clusters \citep{abramowski2012discovery,aharonian2022deep,abeysekara2021hawc}, the analysis cannot rule out contributions from other unresolved sources.

 Figure \ref{fig:final}(b) shows the best-fit spectrum for HAWC J1746-2856 (see Table \ref{tab:sys} for systematic uncertainties and best-fit position) compared to the H.E.S.S. measured spectra of Sgr A$^{*}$ and the radio Arc. Since HAWC cannot resolve these point sources, we conservatively assume that their spectra extends and covers the entire HAWC energy range, which is represented with dashed lines in Figure \ref{fig:final}(b). In Figure \ref{fig:final}(c), we show the significance map obtained after subtracting the estimated excess of the H.E.S.S. point sources from the HAWC data. The predicted event count is calculated by convolving a model consisting of the reported best-fit parameters for the H.E.S.S. sources with the HAWC instrument response function. We also include contours of carbon monosulfide (CS) line emission---integrated from $-200$ km/s to 200 km/s---to show the spatial correlation of the HAWC central excess with the density distribution of the ambient dense gas \citep{tsuboi1999dense}. Thus, the residual shown in Figure \ref{fig:final}(c) is likely emission from the GC ridge diffusion and, in smaller contribution, unresolved sources.
 
In Figure \ref{fig:final}(d), we subtract the flux from the two H.E.S.S. point sources from the HAWC best-fit spectrum (shown separately in Figure \ref{fig:final}(b)). The error band illustrates the combination of HAWC and H.E.S.S. uncertainties in quadrature. We also compare our measurement with the diffuse emission flux points estimated in \citet{2016Natur.531..476H}, where the diffuse emission was derived within an annulus of inner radius 0.15$^{\circ}$---to exclude HESS J1745-290---and outer radius 0.45$^{\circ}$. In that study, a sector ($\sim$66$^{\circ}$) of the annulus is excluded to avoid HESS J1746-285. These excluded regions and the slightly larger radius of the HAWC source (0.46$^{\circ}$) may explain the higher flux detected by HAWC, although both results are still compatible within uncertainties (see Figure \ref{fig:final}(d)). The hard spectrum reported by H.E.S.S. with a photon index of 2.3 \citep{2016Natur.531..476H} is mostly dominated by events with energies below 10 TeV. HAWC is more sensitive at higher energies and measures an index of 2.9. The change in the spectral index occurs at low energies where HAWC is not sensitive enough to probe the cause, given the large zenith angle. However, we find no evidence of significant spectral curvature from 10s of TeV to 114 TeV. Other IACTs have also measured the diffuse emission. However, they use regions with significantly different morphologies: in the studies by H.E.S.S. \citep{abdalla2018characterising} and MAGIC \citep{acciari2020magic} the entire  l$\leq |\pm 1^{\circ}|$ GC region is included, while VERITAS \citet{adams2021veritas} utilized seven circular regions of 0.1° radius outside of the H.E.S.S. \citep{2016Natur.531..476H} annuli.

In summary, we have shown that the measured flux of HAWC J1746-2856 is significantly higher than that of HESS J1745-290 and HESS J1746-285. Therefore, even after excluding their contributions, the spectrum extends beyond 100~TeV.

\newpage
\section{Discussion}\label{sec:disc}

The HAWC detection of photons with energies exceeding 100 TeV further strengthens the hadronic-origin interpretation suggested by \citet{2016Natur.531..476H}, where relativistic protons ($\gtrsim$1~PeV) collide with the surrounding dense ambient gas. 

In the leptonic scenario, the gamma-ray emission comes from the inverse Compton scattering of electrons with energies $E_{e} >100$~TeV. In the GC region, these electrons have a short lifetime, mostly due to synchrotron radiation. Assuming a magnetic field strength of 100~$\mu$G \citep{crocker2010lower}, the cooling time is:
\begin{equation}
t_{\text{cool}} \approx 13 \left(\frac{E_{\text{e}}}{100\,\text{TeV}}\right)^{-1} \left(\frac{B}{100\,\mu\text{G}}\right)^{-2}\,\text{yr}\,,
\end{equation}
corresponding to a maximum distance that the electrons may travel $c\, t_{\text{cool}} = 4\,\text{pc}$, even assuming the extreme case of ballistic movement. Such a distance is significantly smaller than the size of the CMZ, which is hundreds of parsec. Therefore, the HAWC observation strongly disfavors the leptonic scenario. The only way to make such a scenario work would be to have tens of unresolved electron accelerators co-existing in the region.

In the hadronic scenario, although $\pi_{0}$ decay is the dominant cool-down channel \citep{aharonian2009spectrum,longair2010high}, the cooling time is so much larger than the escape time  (by several orders of magnitude) that the proton-cooling effect is negligible \citep{scherer2023modeling}. The escape time of $E_{\text{p}} = 1 $~PeV protons can be roughly estimated as:
\begin{equation}
 t_{\text{escape}} \approx \frac{r^2}{2D} \approx 100 \left(\frac{r}{40\,\text{pc}}\right)^2 \left(\frac{E_{\text{p}}}{1\, \text{PeV}} \right)^{-0.3}\,\text{yr}\,,
 \end{equation}
 where $D\sim 1.2 \times 10^{30}(E_{\text{p}}/100\text{ TeV})^{0.3} \,\text{cm}^2/\text{s}$ \citep{strong2007cosmic} is the diffusion coefficient in the interstellar medium (ISM) and $r\sim 40$ pc is the radius of the diffuse emission region used in \citet{2016Natur.531..476H}. As the magnetic field at the GC is much higher than that of the average ISM \citep{crocker2010lower}, protons are likely confined therein for a longer time. Nonetheless, $t_{\text{escape}}$ is much shorter than the age of the Galaxy, implying that the proton source(s) are either very young or injecting protons into the CMZ in a recent burst. Therefore, the only plausible explanation is that one or more sources quasi-continuously accelerate and inject high-energy protons into the CMZ at rates that exceed the escape time.

Finally, we estimated the gamma-ray luminosity ($L_{\gamma}(E_{\gamma}\geq 10\text{ TeV})=2.24\times10^{34}$ erg/s) by integrating the differential flux of the HAWC central source between 10 and 114 TeV, subtracting the contribution of H.E.S.S. point sources and assuming an 8.5 kpc distance to the GC region. With this result, we calculated the energy density of cosmic-ray protons using our measurement of the gamma-ray flux above 10 TeV to be:
\begin{equation}
    w_{\text{p}}(\geq 10 E_{\gamma})= 1.8 \times 10^{-2}\left(\frac{\eta_{N}}{1.5} \right)^{-1}\left(\frac{L_{\gamma}(E_{\gamma}\geq 10\text{ TeV})}{10^{34}\,\text{erg/s}} \right)\left(\frac{M}{10^{6}M_{\odot}} \right)^{-1}\,\text{eV/cm}^{3}\approx 8.1\times 10^{-3}\,\text{eV/cm}^{3}   \,,
    \label{eq:EnergyDensity}
\end{equation}
where the CS total mass of the gas ($5\times 10^{6}M_{\odot}$) is the sum of CS mass in the three H.E.S.S. annuli that are roughly coincident with the HAWC region \citep{2016Natur.531..476H} and $\eta_{N}=1.5$ considers the existence of nuclei heavier than hydrogen in cosmic rays and the interstellar matter. This energy density obtained for $>$100~TeV protons is larger than the $1\times 10^{-3}$~eV/cm$^{3}$ local measurement by the Alpha Magnetic Spectrometer (AMS; \citealp{aguilar2015precision,abeysekara2021hawc}).   
Additionally, we calculate the total energy budget of protons with energies $>$100~TeV: 
\begin{equation}
    W_{p}\approx L_{\gamma}(E_{\gamma}\geq 10\text{ TeV}) \, t_{\text{pp}} \approx 3.53  \times 10^{49} n^{-1}\,\text{erg}\,,
\end{equation}
where $t_{\text{pp}}\approx5\times 10^{7}n^{-1}$ yr is the cooling time for proton-proton (pp) interactions assuming the relative velocity of the interacting protons to be equivalent to the speed of light ($c$) and an ambient gas density of $n$, in units of cm$^{-3}$.
We estimated the cosmic-ray energy density from H.E.S.S. measurements using the diffuse region shown in Figure~\ref{fig:final}. By integrating the protons with energies between 100~TeV and 1140~TeV, we found the integral cosmic-ray density to be $\approx 2.1 \times 10^{49} n^{-1}\,\text{erg}$, which is compatible with HAWC's results. Our interpretation is consistent with the steady proton source scenario suggested by \citet{2016Natur.531..476H}. Therefore, we attribute the UHE gamma rays to the freshly accelerated proton cosmic rays from the local accelerators within the GC region, which continuously inject protons with PeV energies.

\section{Conclusions}\label{section:conc}

We report the first detection of $>$100~TeV gamma rays from the GC region with a number of nearly 100 events. This HAWC result extends the highest energy reported from the GC by the IACTs by more than a factor of two. The best-fit model for seven years of HAWC data from the GC is a point source with a simple-power-law spectrum ($\mathrm{d}N/\mathrm{d}E=\phi(E/26 \,\text{TeV})^{\gamma}$), where $\gamma=-2.88 \pm 0.15_{\text{stat}} - 0.1_{\text{sys}} $ and $\phi=1.5 \times 10^{-15}$ (TeV cm$^{2}$s)$^{-1}$ $\pm\, 0.3_{\text{stat}}\,^{+0.08_{\text{sys}}}_{-0.13_{\text{sys}}}$, with no signs of a cutoff. After subtracting the small contribution of HESS J1745-290 and HESS J1746-285 from the HAWC best-fit spectrum, the remaining flux---likely from the Galactic ridge diffuse emission---maintains the power-law shape, extending to at least 114 TeV. Extending the power-law spectrum to these energies reveals a PeVatron at the GC, as first suggested by \citet{2016Natur.531..476H}, with photons up to $\sim$30~TeV. Although our analysis does not resolve the object accelerating protons to PeV energies, we can confirm the existence of a PeVatron at the GC. Additionally, we discuss the possible origin of such high-energy gamma rays---using model-independent arguments---and conclude that the hadronic mechanism and quasi-continuous injection scenarios are preferred. Moreover, we calculate the gamma-ray luminosity of the PeVatron and find that the cosmic-ray energy density is above the average, which clearly suggests the presence of freshly accelerated 0.1--1~PeV protons in the GC region. Finally, we show that the total energy budget of protons with energies $>$100~TeV calculated with HAWC data is compatible with H.E.S.S. measurements.

  Several specific sites of proton acceleration have been proposed within the HAWC J1746-2856 emission region, in particular near the vicinity of Sgr A$^{*}$ \citep{2016Natur.531..476H} and within the compact star clusters, the Arches and Quintuplet clusters \citep{aharonian2019massive}, which we did not resolve in this analysis. Recently, there has been progress in modeling the CMZ with more realistic cosmic-ray dynamics in agreement with existing data \citep{scherer2023modeling}. The next generation of experiments, such as the Cherenkov Telescope Array (CTA; \citealp{consortium2019cherenkov}) and the Southern Wide-field Gamma-ray Observatory (SWGO; \citealp{2019arXiv190208429A}), could better differentiate and constrain these models with improved gamma-ray observations. 
 
\begin{acknowledgments}
We acknowledge the support from: the US National Science Foundation (NSF); the US Department of Energy Office of High-Energy Physics; the Laboratory Directed Research and Development (LDRD) program of Los Alamos National Laboratory; Consejo Nacional de Ciencia y Tecnolog\'{i}a (CONACyT), M\'{e}xico, grants 271051, 232656, 260378, 179588, 254964, 258865, 243290, 132197, A1-S-46288, A1-S-22784, CF-2023-I-645, c\'{a}tedras 873, 1563, 341, 323, Red HAWC, M\'{e}xico; DGAPA-UNAM grants IG101323, IN111716-3, IN111419, IA102019, IN106521, IN114924, IN110521 , IN102223; VIEP-BUAP; PIFI 2012, 2013, PROFOCIE 2014, 2015; the University of Wisconsin Alumni Research Foundation; the Institute of Geophysics, Planetary Physics, and Signatures at Los Alamos National Laboratory; Polish Science Centre grant, DEC-2017/27/B/ST9/02272; Coordinaci\'{o}n de la Investigaci\'{o}n Cient\'{i}fica de la Universidad Michoacana; Royal Society - Newton Advanced Fellowship 180385; Generalitat Valenciana, grant CIDEGENT/2018/034; The Program Management Unit for Human Resources \& Institutional Development, Research and Innovation, NXPO (grant number B16F630069); Coordinaci\'{o}n General Acad\'{e}mica e Innovaci\'{o}n (CGAI-UdeG), PRODEP-SEP UDG-CA-499; National Research Foundation of Korea RS-2023-00280210; Institute of Cosmic Ray Research (ICRR), University of Tokyo. H.F. acknowledges support by NASA under award number 80GSFC21M0002. We also acknowledge the significant contributions over many years of Stefan Westerhoff, Gaurang Yodh and Arnulfo Zepeda Dom\'inguez, all deceased members of the HAWC collaboration. Thanks to Scott Delay, Luciano D\'{i}az and Eduardo Murrieta for technical support.
\end{acknowledgments}
\noindent S. Yun-Cárcamo analyzed the data, performed the maximum likelihood analysis and prepared the original manuscript. S. Yun-Cárcamo and D. Huang carried out the discussion section calculations. R. Babu, K. L. Fan, and D. Huang helped with analysis tools and interpretation of results. The full HAWC Collaboration has contributed through the construction, calibration, and operation of the detector, the development and maintenance of reconstruction and analysis software, and vetting of the analysis presented in this manuscript. All authors have reviewed, discussed, and commented on the results and the manuscript.

\vspace{5mm}





\bibliography{sample631}{}
\bibliographystyle{aasjournal}



\end{document}